\documentclass{aa}
\usepackage{txfonts}

\usepackage{booktabs}       % professional-quality tables
\usepackage{nicefrac}  % Tilted fractions via `nicefrac`
\usepackage[separate-uncertainty=true]{siunitx}  % Give numbers units (with proper spacing)

\usepackage{placeins}  % place float barriers
\usepackage{graphicx}  % Insert images (also works with spaces in filename)
\usepackage{natbib}  % A&A compatible citation
\usepackage{hyperref}  % Clickable references
\hypersetup{colorlinks=true, allcolors=blue}  % No green boxes for links
\usepackage[nameinlink]{cleveref}  % Pass multiple or even a range as reference, e.g. Figure 1-5
\usepackage{subcaption}  % Customize sub-captions
\bibpunct{(}{)}{;}{a}{}{,} % to follow the A&A style

\begin{document}

\title{The ``C'': The large Chameleon-Musca-Coalsack cloud}
% \title{The Chameleon, Musca, and Coalsack Cloud: a C-ring}
% \title{The Chameleon ``C'': a well-defined half-ring containing the Chameleon, Musca, and Coalsack cloud complexes}  % Iterative Charted Extinction Ball
% \subtitle{}
% \titlerunning{The ``C''}

\author{Gordian~Edenhofer\inst{1,2,3}\fnmsep\thanks{\email{edh@mpa-garching.mpg.de}}
   \and
   João~Alves\inst{3}
   \and
   Catherine~Zucker\inst{4}
   \and
   Laura~Posch\inst{3}
   \and
   Torsten~A.~Enßlin\inst{1,2,5}
}

\institute{
   Max Planck Institute for Astrophysics,
   Karl-Schwarzschild-Straße 1, 85748 Garching bei München, Germany
   \and
   Ludwig Maximilian University of Munich,
   Geschwister-Scholl-Platz 1, 80539 München, Germany
   \and
   Institute for Astronomy,
   University of Vienna, Türkenschanzstrasse 17, 1180 Vienna, Austria
   \and
   Center for Astrophysics $\vert$ Harvard \& Smithsonian,
   60 Garden St., Cambridge, MA 02138, USA
   \and
   Excellence Cluster ORIGINS, Boltzmannstraße 2, 85748 Garching bei München, Germany
}

\date{}

% \abstract{}{}{}{}{}
% 5 {} token are mandatory
\abstract{
Recent advancements in 3D dust mapping have transformed our understanding of the Milky Way's local interstellar medium, enabling us to explore its structure in three spatial dimensions for the first time.
In this Letter, we use the most recent 3D dust map by Edenhofer et al. to study the well-known Chameleon, Musca, and Coalsack cloud complexes, located about \SI{200}{pc} from the Sun.
We find that these three complexes are not isolated but rather connect to form a surprisingly well-defined half-ring, constituting a single C-shaped cloud with a radius of about \SI{50}{pc}, a thickness of about \SI{45}{pc}, and a total mass of about \SI{5e4}{M_{\odot}}, or \SI{9e4}{M_{\odot}} if including everything in the vicinity of the C-shaped cloud.
Despite the absence of an evident feedback source at its center, the dynamics of young stellar clusters associated with the C structure suggest that a single supernova explosion about \SIrange{4}{10}{Myr} ago likely shaped this structure.
Our findings support a single origin story for these cloud complexes, suggesting that they were formed by feedback-driven gas compression, and offer new insights into the processes that govern the birth of star-forming clouds in feedback-dominated regions, such as the Scorpius–Centaurus association.
}

\maketitle

\section{Introduction}
\label{sec:introduction}

For over a hundred years, our understanding of the interstellar medium (ISM) has been limited to 2D projections.
Analysis of 2D projections, while a mainstay in traditional studies of the ISM, can create a deceptive picture of the true and complex 3D structures of the ISM.
For example, atomic and molecular clouds can appear as distinct or overlapping entities, which obscures their true spatial relationships.
This illusion is especially problematic when clouds exhibit similar radial velocities, making it nearly impossible to distinguish their boundaries and depths using traditional observational methods.

The nearby molecular clouds of Chameleon, Musca, and Coalsack -- long suspected of being physically associated with one another -- are a potential example of this type of confusion \citep{Corradi1997,Corradi2004CoalsackChamaeleonMusca}.
However, their seemingly distinct appearances in the 2D plane of the sky has complicated any firm conclusions regarding a possible physical association.
This projected view is potentially hiding diffuse connections that could shed new light on their origin and evolution.

In recent years, the proliferation of 3D dust maps has begun a revolution in the field \citep[e.g.,][]{Edenhofer2023,Leike2019,Vergely2022,Lallement2022,Lallement2019,Lallement2018,Capitanio2017,Babusiaux2020,Hottier2020,Green2019,Green2017,Leike2022,Chen2019,Rezaei2022,Rezaei2020,Rezaei2018,Rezaei2017,Dharmawardena2022}.
These new maps capitalize on the astrometric data from the ESA \textit{Gaia} satellite \citep{GaiaCollaboration2022} and have emerged as powerful tools for unraveling the complex volume density distribution of gas within the local Milky Way, providing unprecedented insights into the structure and dynamics of molecular clouds from kiloparsec down to parsec scales \citep{Zucker2023-pp, Zucker2018-up,Zucker2019,Zucker2020-gj,Alves2020-um,Konietzka2024,Zucker2021-xq,Kuhn2022-nv,Posch2023-in}.

Chameleon, Musca, and Coalsack are among the closest molecular clouds to Earth, at a distance of about \SI{200}{pc} \citep[see, e.g.,][]{Zucker2021,Corradi2004CoalsackChamaeleonMusca}.
Chameleon has been extensively studied for its active star formation \citep[e.g.,][]{Luhman2008-md,Belloche2011,Persi2003,Tsitali2015} and Musca for its simple filamentary structure and subsonic nature \citep[e.g.,][]{Mizuno1998-sd,Hacar2016-yv,Kainulainen2016-nc}.
Coalsack, despite being one of the few dark clouds visible to the naked eye, is one of the least studied nearby clouds, probably because it is not star-forming and is seen against the complicated background of the Galactic plane \citep[e.g.,][]{Nyman2008-cy,Beuther2011-ov}.
So far, all three clouds have been treated as separate entities, and conflicting origin stories have been proposed for each.
For example, Musca, is hypothesized to have been shaped by magnetic fields, by the dissipation of supersonic turbulence, by a cloud--cloud collision, or a combination of these processes \citep{Cox2016-aq,Tritsis2018-mq,Tritsis2022TheMuscaMolecularCloud,Bonne2020-we, Bonne2020-gm,Yahia2021-pd,Kaminsky2023-ik}.

In this Letter we study the Chameleon, Musca, and Coalsack region from a new perspective, namely, in full spatial 3D.
Using the 3D dust map from \citeauthor{Edenhofer2023}, we characterize the topology and mass of this region and analyze the relationship between the three molecular clouds.
In addition, we analyze the dynamics of young stellar clusters (YSO clusters) embedded in Chameleon and Coalsack as a proxy for the large-scale dynamics of the gas.

\section{Methods and results}
\label{sec:methods_and_results}

We used the 3D reconstruction of the interstellar dust distribution presented in \citet{Edenhofer2023} throughout our analysis.
The reconstruction is based on the catalog described in \citet{Zhang2023}, which in turn is based on the ESA \textit{Gaia} DR3 BP/RP spectra~\citep{GaiaCollaboration2022}.
Using the stellar extinction and distances in the catalog, \citeauthor{Edenhofer2023} reconstructed the 3D distribution of interstellar dust.
The map extends to \SI{1.25}{kpc} from the Sun and achieves \SI{14}{'} angular resolution and parsec-scale distance resolution.
\citeauthor{Edenhofer2023} discretized the modeled 3D space into logarithmically spaced distance voxels and equal-area plane-of-sky voxels.
In this discretized space, they represent the logarithm of the differential dust extinction using a Gaussian process model \citep{Edenhofer2022}.
The inference uses the software package NIFTy \citep{Edenhofer2024NIFTyRE}.

\citet{Edenhofer2023} have created one of the highest resolution 3D dust maps of the local Milky Way.
Unlike other 3D dust maps in the literature, it focuses on a relatively small volume \citep[cf.][]{Green2019}, but resolves the structures within this volume with high (parsec-scale) resolution, comparable to~\citet{Leike2020}.
Compared to \citet{Leike2020}, it reconstructs a much larger volume and yields a higher dynamic range thanks to the new \textit{Gaia} data.
The methodology extends that from \citet{Leike2020} and employs its non-parametrically inferred correlation kernel.
Furthermore, the model strictly enforces physical constraints, such as positive definiteness of differential interstellar dust densities, and rigorously incorporates the uncertainties on the distances to stars.  % instead of employing ad hoc uncertainty corrections
The inference algorithm used has been extensively proven in real-world applications \citep[see][]{Galan2024ElGordo,Arras2022,Leike2019,Leike2020,Mertsch2023,Roth2023DirectionDependentCalibration,Hutschenreuter2023,Tsouros2023,Roth2023FastCadenceHighContrastImaging,Hutschenreuter2022} and is provably exact in the linear regime \citep{Knollmueller2019}.

\subsection{Topology}

Using the 3D map of the distribution of interstellar dust from \citet{Edenhofer2023}, we analyzed the region around Chameleon, Musca, and Coalsack.
For convenience, we interpolated the 3D map with irregularly spaced voxels to a regular Cartesian and a regular spherical grid using the scripts provided by the authors online\footnotemark{} and distributed as part of the Python package \texttt{dustmaps} \citep{Green2018Dustmaps}.
We find that all three molecular clouds are embedded in a large 3D structure that forms a C-shaped half-ring.
The structure lies at the edge of the Local Bubble \citep{Zucker2022StarFromationNearTheSun,ONeill2024LocalChimney,Pelgrims2020LocalBubble}, close to the Scorpius-Centaurus association.
Due to obscuration and projection effects, the peculiar C shape is only revealed using 3D reconstructions of interstellar dust.
In an ordinary 2D plane-of-sky projection, the diffuse bridges connecting the molecular clouds become indistinguishable from faint dust extinction at larger distances.
\footnotetext{%
    \url{https://zenodo.org/doi/10.5281/zenodo.8187942}
}

\begin{figure}[!htbp]
   \centering
   \subcaptionbox{}{%
      \includegraphics[width=0.95\hsize,keepaspectratio]{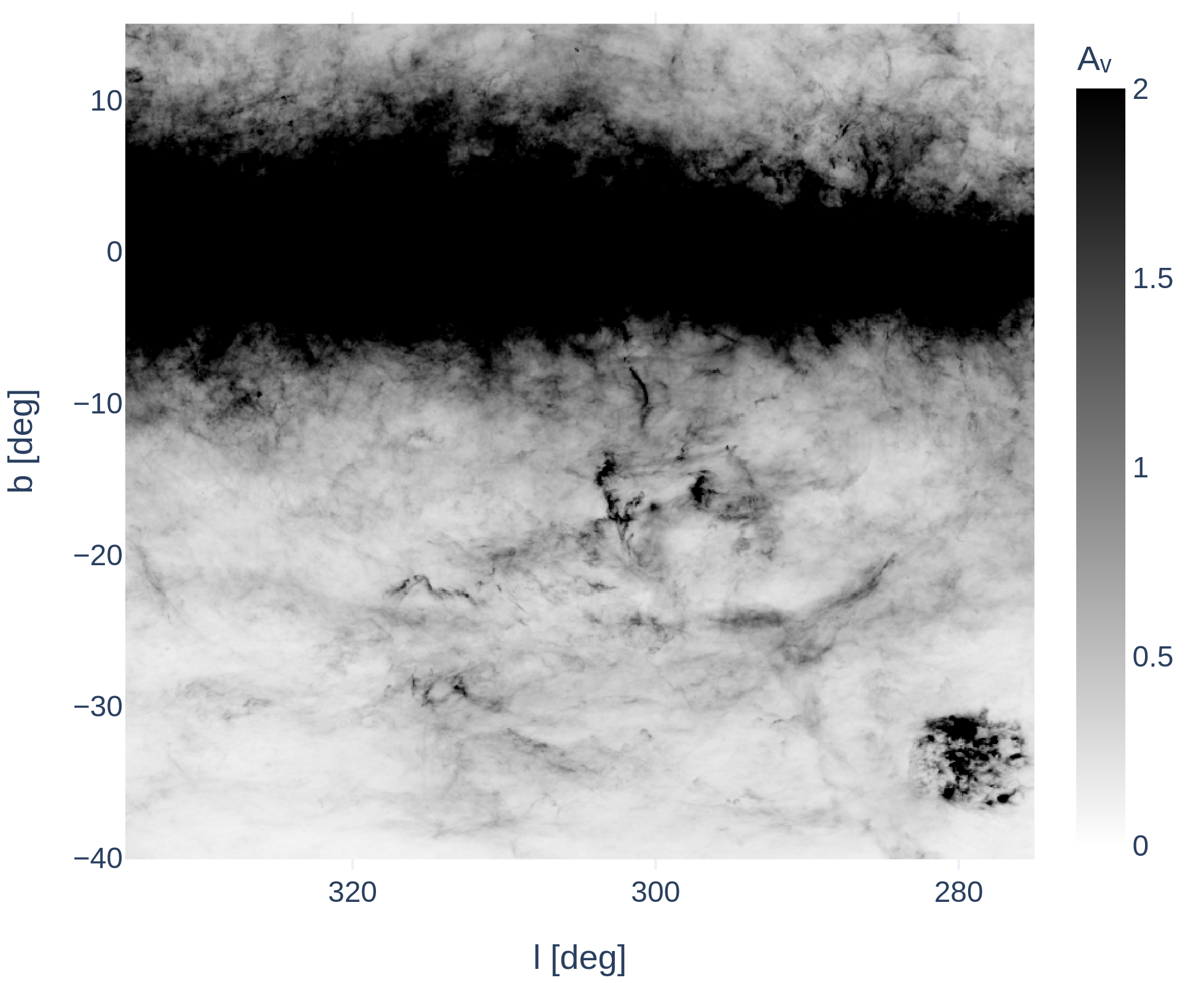}
   }
   \subcaptionbox{}{%
      \includegraphics[width=0.95\hsize,keepaspectratio]{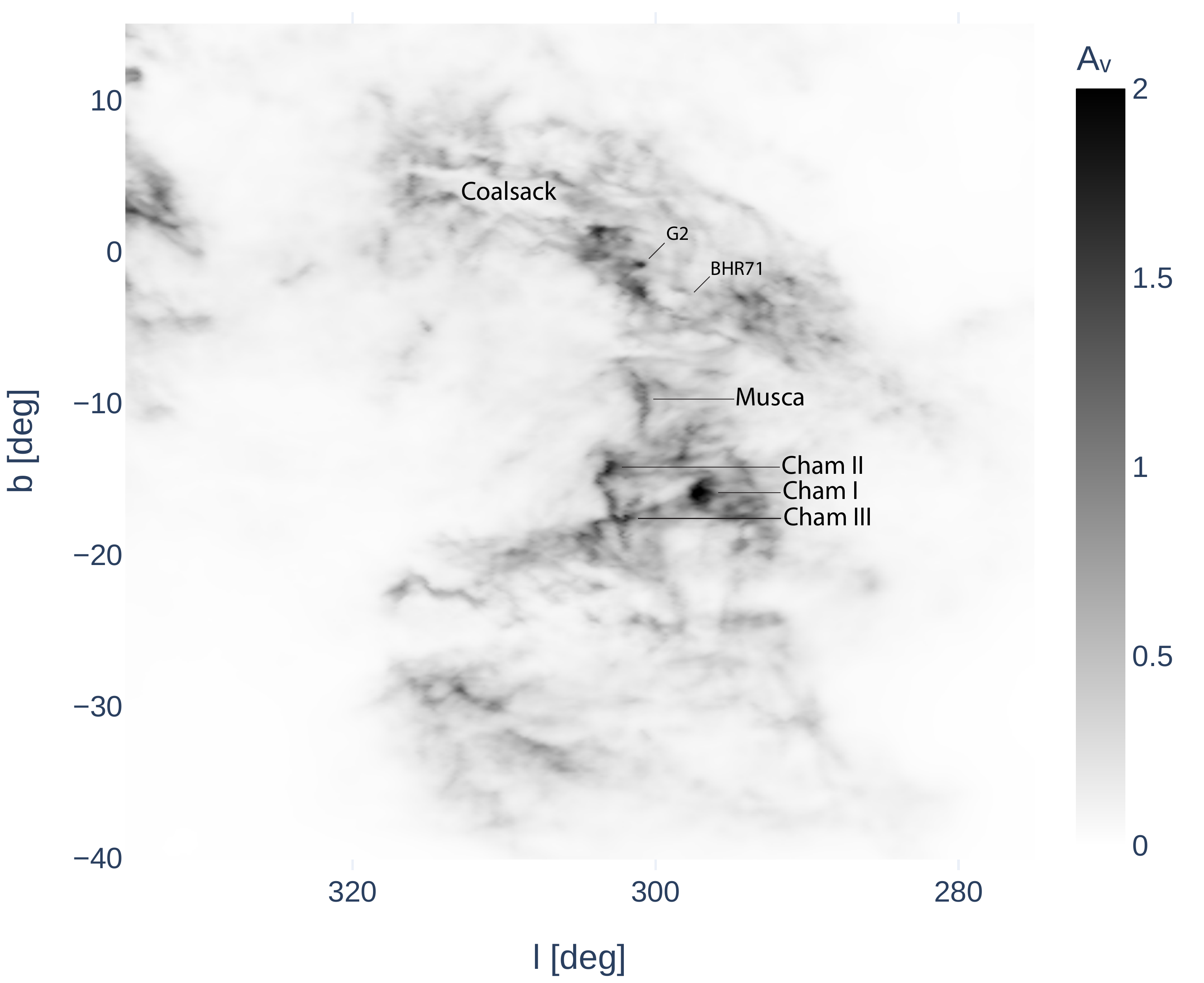}
   }
   \caption{%
        Plane-of-sky region toward the C of the posterior mean of \cite{Edenhofer2023}.
        Panel (a): \textit{Planck}~2013 extragalactic E(B-V) extinction toward the C converted to $A_\mathrm{V}$ via $A_\mathrm{V} = 3.1 \cdot \mathrm{E(B-V)}$.
        Panel (b): Visual extinction, $A_\mathrm{V}$, between \SI{165}{pc} and \SI{220}{pc} toward the C-shaped structure in the 3D interstellar dust map.
        Spurs of Lupus are seen on the left edge.
        Both panels display the extinction in units of magnitude, and the colorbars are linear but clipped at an extinction of $A_\mathrm{V} = \SI{2}{mag}$.
        \label{fig:edenhofer23_lb}
   }
\end{figure}

\autoref{fig:edenhofer23_lb} shows the extinction within the region over the range \SIrange{335}{275}{\degree} in Galactic longitude and \SIrange{-40}{15}{\degree} in Galactic latitude.
The first panel shows the \textit{Planck}~2013 extragalactic $\mathrm{E(B-V)}$ extinction of interstellar dust \citep{Planck2013} converted to $A_\mathrm{V}$ via $A_\mathrm{V} = 3.1 \cdot \mathrm{E(B-V)}$.
The C-shaped structure is almost completely obscured.
The second panel shows the 3D interstellar dust map integrated over a distance of \SI{55}{pc} (from \SIrange{165}{220}{pc}).
Here, the C-shaped structure appears as a single coherent structure that hosts Chameleon, Musca, and Coalsack.
The individual dense molecular clouds are connected through faint lanes of interstellar dust extinction.
Thanks to the 3D distance selection, the figure shows this region free of confusion stemming from extinction at farther distances.

\begin{figure}[!htbp]
   \centering
   \includegraphics[height=0.28\textheight,keepaspectratio]{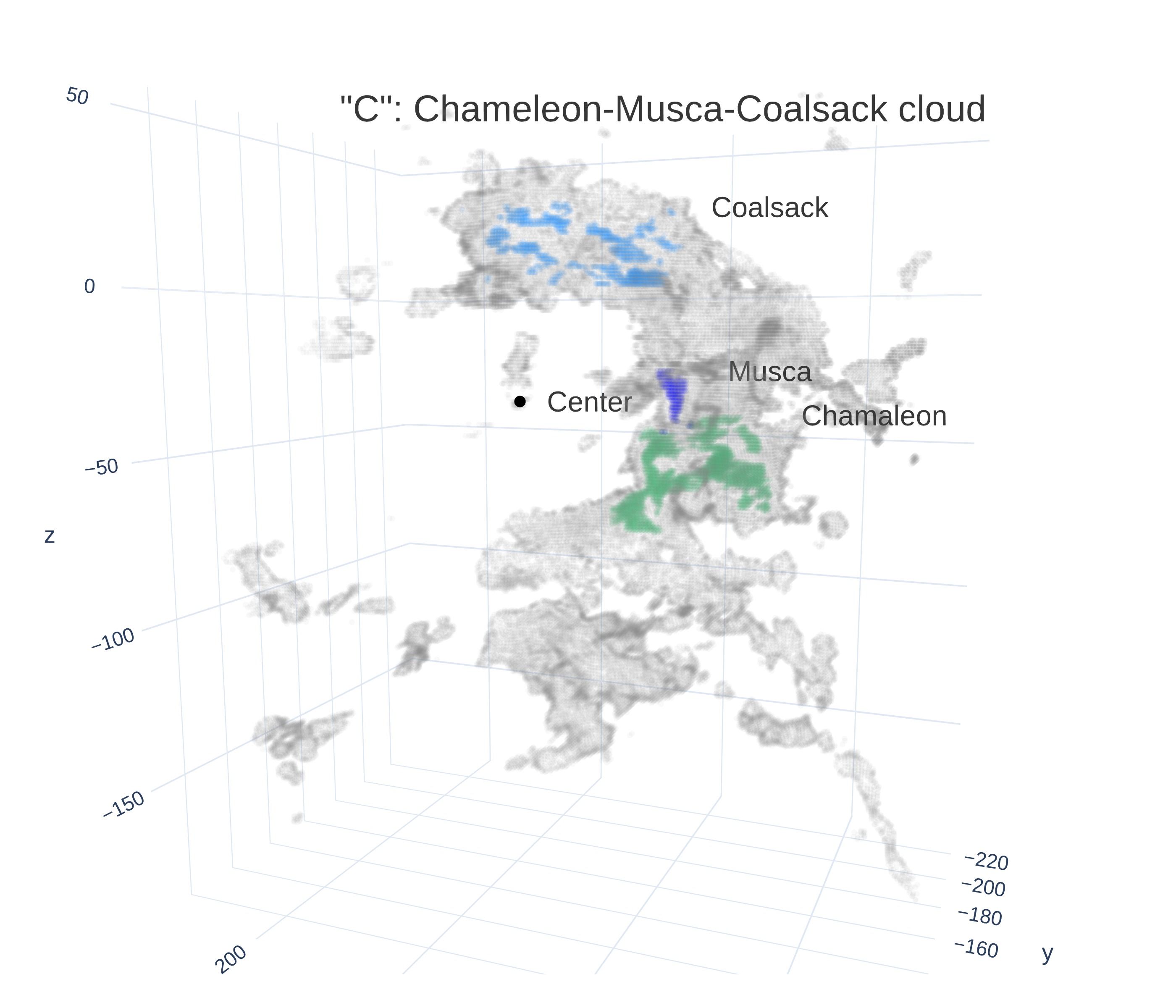}
   \caption{%
      3D view of the isodensity surface of the posterior mean of the 3D dust map of \citeauthor{Edenhofer2023} within the Cartesian selection given in \Cref{tab:parameters}, showing the C.
      High-density structures ($n_\mathrm{H} \geq \SI{23.3}{cm^{-3}}$; the $10,000$ highest-density points of a \SI{1}{pc} Cartesian interpolation of the C) toward Chameleon are color-coded in green ($\SI{290}{\degree} \leq l \leq \SI{306}{\degree}$ and $\SI{-21}{\degree} \leq b \leq \SI{-12}{\degree}$), toward Musca in violet ($\SI{300}{\degree} \leq l \leq \SI{302.5}{\degree}$ and $\SI{-13}{\degree} \leq b \leq \SI{-7.25}{\degree}$), and toward Coalsack in blue ($\SI{300}{\degree} \leq l \leq \SI{318}{\degree}$ and $\SI{0.5}{\degree} \leq b \leq \SI{7.5}{\degree}$).
      An interactive version of this figure is available at \url{https://faun.rc.fas.harvard.edu/gedenhofer/perm/C/C_Chameleon_Musca_Coalsack_cloud.html}.
      \label{fig:edenhofer23_isosurface}
   }
\end{figure}

With 3D interstellar dust maps, we are not restricted to plane-of-sky projections and can instead explore the distribution of interstellar dust in full spatial 3D.
In \autoref{fig:edenhofer23_isosurface} we show the isodensity surface in full spatial 3D with a total hydrogen nucleus density of approximately $n_\mathrm{H}=\SI{4.5}{cm^{-3}}$ (cf. \citealt{Zucker2021}; 0.98 quantile of the 3D dust density in the selected volume; see \citealt{ONeill2024LocalChimney} on how to convert the units of the 3D dust density to $n_\mathrm{H}$).
The outline traces the C-shaped spine observed in \autoref{fig:edenhofer23_lb}.
The isodensity surface fully envelops the half-ring that includes the Chameleons, Musca, and Coalsack molecular clouds.
We term the C-shaped structure the ``C.''

Based on its suggestive half-ring structure, we defined a center for the C.
We defined it to lie below Coalsack in Galactic X and Y and slightly above the Galactic Z of the two Chameleons.
Specifically, we set the center to be at Galactic X=\SI{138}{pc}, Y=\SI{-140}{pc}, and Z=\SI{-33}{pc} (Galactic $l=\SI{314.6}{\degree}$, $b=\SI{-9.5}{\degree}$, and $d=\SI{199.3}{pc}$; see the interactive \Cref{fig:edenhofer23_isosurface}).

Relative to the center location, we defined a radius using the volume filling fraction of interstellar dust.
We find that the volume filling fraction reaches a maximum at a distance of \SI{48}{pc}, and we assume this to be the radius of the C.
This estimate is robust with respect to different posterior samples of the dust distribution from \citet{Edenhofer2023}.
The full width at half maximum (FWHM) of the peak is \SI{47\pm2}{pc}, and most of the dust lies within \SI{80}{pc} of the center (the ceiling of the more distant edge of the FWHM).
% We refer to \autoref{appx:Volume_filling_fraction_and_mass} for details on the volume filling fraction.

We find that the C is almost perfectly parallel to the Galactic Z axis.
Given its proximity to the Galactic disk, this makes it perpendicular to the line of sight.
To quantify the inclination, we selected the 10,000 highest-density points of a \SI{1}{pc} interpolation of the 3D dust map within $20 \leq X \leq 230$~pc, $-240 \leq Y \leq -90$~pc, and $-185 \leq Z \leq 50$~pc and decomposed them with a singular value decomposition into a plane.
We find that the Galactic Z axis and the plane's normal vector form an \SI{86}{\degree} angle (two-thirds of the samples are within $\pm\SI{1}{\degree}$) and the inclination between the line of sight toward the center and the fitted plane to be \SI{72}{\degree} (two-thirds of the samples are within $\pm \SI{1}{\degree}$).
\autoref{tab:parameters} summarizes the key properties of the C.

% In [26]: v = fwhm_upr; m = np.mean(v); s = np.std(v); n_o = np.sum((np.array(v) <= m - s) | (np.array(v) >= m + s
%     ...: )); q = np.quantile(v, (2/12, 10/12))
%     ...: print(f"{v}\n{m} || {s} || {n_o} || {q}")
% [75.5, 75.5, 75.5, 76.5, 76.5, 76.5, 75.5, 76.5, 76.5, 75.5, 75.5, 75.5]
% 75.91666666666667 || 0.4930066485916347 || 5 || [75.5 76.5]
% In [27]: v = fwhm_lwr; m = np.mean(v); s = np.std(v); n_o = np.sum((np.array(v) <= m - s) | (np.array(v) >= m + s
%     ...: )); q = np.quantile(v, (2/12, 10/12))
%     ...: print(f"{v}\n{m} || {s} || {n_o} || {q}")
% [31.5, 30.5, 30.5, 29.5, 28.5, 29.5, 28.5, 28.5, 29.5, 29.5, 29.5, 30.5]
% 29.666666666666668 || 0.8975274678557508 || 4 || [28.5 30.5]
% In [28]: v = mass; m = np.mean(v); s = np.std(v); n_o = np.sum((np.array(v) <= m - s) | (np.array(v) >= m + s));
%     ...: q = np.quantile(v, (2/12, 10/12))
%     ...: print(f"{v}\n{m} || {s} || {n_o} || {q}")
% [93.6, 92.7, 93.1, 93.5, 93.4, 93.3, 94.0, 93.9, 93.4, 93.4, 93.1, 93.2]
% 93.38333333333333 || 0.33870669054836033 || 3 || [93.1  93.65]
\begin{table*}[!htbp]
   \caption{%
      Key properties of the C.
   }
   \label{tab:parameters}
   \centering
\begin{tabular}{@{}cccccccc@{}}
\toprule
 &
  \begin{tabular}[c]{@{}c@{}}Cartesian\\ Position\end{tabular} &
  \begin{tabular}[c]{@{}c@{}}Spherical\\ Position\end{tabular} &
  Radius &
  FWHM &
  Inclination &
  Mass &
  \\ \midrule
C &
  \begin{tabular}[c]{@{}c@{}}$\SI{20}{pc} \leq X \leq \SI{230}{pc}$\\ $\SI{-240}{pc} \leq Y \leq \SI{-90}{pc}$\\ $\SI{-185}{pc} \leq Z \leq \SI{50}{pc}$\end{tabular} &
  \begin{tabular}[c]{@{}c@{}}$\SI{275}{\degree} \leq l \leq \SI{335}{\degree}$\\ $\SI{-40}{\degree} \leq b \leq \SI{15}{\degree}$\\ $\SI{165}{pc} \leq d \leq \SI{220}{pc}$\end{tabular} &
  \begin{tabular}[c]{@{}c@{}}\SI{48}{pc}\\ (all samples)\end{tabular} &
  {\begin{tabular}[c]{@{}c@{}c@{}} \SI{29.5 \pm 1}{pc} \\ -- \\ \SI{76\pm0.5}{pc}\end{tabular}} &
  \SI{72\pm1}{\degree} &
  \begin{tabular}[c]{@{}c@{}}$(50.5\pm0.3)\times10^{3}\,\si{M_{\odot}}$\\ in isodensity surface\end{tabular} &
  \\
Center &
  \begin{tabular}[c]{@{}c@{}}X=\SI{137}{pc}\\ Y=\SI{-140}{pc}\\ Z=\SI{-33}{pc}\end{tabular} &
  \begin{tabular}[c]{@{}c@{}}l=\SI{314.6}{\degree}\\ b=\SI{-9.5}{\degree}\\ d=\SI{199.3}{pc}\end{tabular} &
   &
   &
   &
   \begin{tabular}[c]{@{}c@{}}$(93.4\pm0.3)\times10^{3}\,\si{M_{\odot}}$\\ in vicinity\end{tabular} &
   \\ \cmidrule(r){1-8}
\end{tabular}
\tablefoot{
    The specified uncertainties encompass at least two-thirds of the 3D dust map posterior samples.
    The specified positions capture the core features of the C while including as little as possible from other molecular clouds.
    The spherical and Cartesian cuts are not equivalent as for each of them we tried to minimize projection and obscuration effects from molecular clouds not affiliated with the C (cf.~\Cref{fig:edenhofer23_lb,fig:edenhofer23_xyz}).
}
\end{table*}

To estimate the mass, we again utilized the 3D interstellar dust density.
We integrated the mass from the center out to a distance of \SI{80}{pc} or the edge of the Cartesian selection given in \Cref{tab:parameters}, whichever is lower, and find the mass of everything within the vicinity of the C to be $(9.34 \pm 0.03) \times 10^{4}\,\si{M_{\odot}}$ (with the statistical uncertainty covering the 0.16 to 0.84 quantile of the interstellar 3D dust map).
This mass estimate is roughly between the total mass of everything above a density of $n_\mathrm{H}=\SI{1}{cm^{-3}}$ and $n_\mathrm{H}=\SI{2}{cm^{-3}}$ within the Cartesian selection given in \Cref{tab:parameters}, $(1.12 \pm 0.01) \times 10^5\,\si{M_{\odot}}$ and $(8.00 \pm 0.08)\times10^4\,\si{M_{\odot}}$, respectively.
The mass of everything within the isosurface shown in \Cref{fig:edenhofer23_isosurface} (i.e., above approximately $n_\mathrm{H}=\SI{4.5}{cm^{-3}}$) is $(5.05 \pm 0.03) \times 10^{4}\,\si{M_{\odot}}$.
Details on how we computed the mass are provided in \Cref{appx:mass}.

\subsection{Dynamics}
\label{sec:dynamics}

To investigate the dynamics of the peculiar C-shaped structure, we studied YSO clusters embedded in the C.
We used the \citet{Ratzenboeck2023} and \citet{Hunt2023} catalogs, applying the following cuts to the \citeauthor{Hunt2023} catalog: $\mathrm{logAge} \leq \log_{10}({\SI{30}{Myr}})$, $\text{astrometric S/N (SNR)} \geq 5$, and $50\mathrm{th}$ percentile of color magnitude diagram (CMD) class $\geq 0.5$ \citep[cf.][]{Hunt2023}.
Nested inside the C, we find three YSO clusters: Centaurus-Far \citep[HSC 2630 in][]{Hunt2023}, Chameleon~I \citep{Hunt2023,Ratzenboeck2023}, and Chameleon~II \citep{Ratzenboeck2023}.
In the following, we use the velocities given in \cite{Ratzenboeck2023}.

The two Chameleons are embedded in the dense centers of the C, while Centaurus-Far lies at its edge.
In contrast to the two Chameleon clusters, Centaurus-Far is strongly extended \citep[see the updated online figure\footnotemark{} from][]{Ratzenboeck2023ScoCen} and encompasses both mostly dust-free regions and dense parts of Coalsack.
Even though Centaurus-Far is not exclusively in the densest regions of the C, as a YSO cluster it likely still traces the overall gas motion of this region.
\footnotetext{\url{https://homepage.univie.ac.at/sebastian.ratzenboeck/wp-content/uploads/2024/04/scocen_ages-2.html}}

Relative to the local standard of rest \citep{Schoenrich2010LSR}, we find that the two Chameleon clusters are moving in the negative Galactic Z direction more strongly than Centaurus-Far.
This implies that the C is expanding at least in the Galactic Z direction.
We find no significant relative motion in the Galactic X and Y directions.
If we assign our center point a hypothetical velocity of $\nicefrac{1}{2} \cdot \left( v_\mathrm{CenFar} + \nicefrac{1}{2} \cdot \left( v_\mathrm{ChamI} + v_\mathrm{ChamII} \right) \right),$ with $v_\square$ the velocities of Centaurus-Far and the two Chameleons, respectively, we find that Centaurus-Far moves up and the two Chameleons move down relative to the center.
Their velocity relative to the center is \SI{3}{km/s} for Centaurus-Far and \SI{4}{km/s} and \SI{2}{km/s} for the two Chameleons.
In Galactic Z, the expansion is \SIrange{1.9}{2.7}{km/s}.
We validate the relative velocities in \Cref{appx:validation_of_dynamics} and find that the velocity uncertainty is on the order of the relative velocities, making this finding noteworthy but highly uncertain.
We note that other tracers for the velocity, such as HI or CO, are unavailable since, in addition to obscuration and confusion, the inclination of the C results in its expansion being perpendicular to the line of sight.

As the YSOs clusters are embedded in the C, we propose that not only do the cluster move away from the center at an average velocity of \SIrange{2}{3}{km/s}, but the whole C expands at an average velocity of \SIrange{2}{3}{km/s}.
Moving the mass in the vicinity of the C at this speed requires a significant amount of energy.
Specifically, moving \SI{9e4}{M_{\odot}} at \SIrange{2}{3}{km/s} requires an energy input of \SIrange{4e48}{8e48}{erg}.
This energy input is $\approx 1\%$ of the total energy release of a supernova feedback event and on the order of the expected kinetic energy input of a supernova \citep{Kim2015-dj}.
% COMMENT: Neglecting acceleration and uncertainties on the velocities, the half-ring would thus have needed 16 Myr to expand from a hypothetical point to its current shape.

\subsection{Age}

We analyzed the relative position of the clusters in the past by propagating their current position and velocity back in time using the software package \texttt{galpy} and the \texttt{MWPotential2014} Galactic potential \citep{Bovy2015}.
% (aconsts.G * (50_000 * aconsts.M_sun) / (50 * u.pc)**2 * 10e+6 * u.yr).to(u.km/u.s)
In doing so, we neglected other gravitational forces such as the cluster's own gravitational potential, gravitational forces due to nearby molecular clouds, and any other acceleration not due to the Milk Way's potential.
% (\SI{50e+3}{\Msun} at a distance of \SI{50}{pc} over \SI{10}{Myr} would contribute approximately \SI{1}{km/s})
We find that the clusters have been approaching each other in the past \SIrange{4}{10}{Myr}.
Relative to the moving center with the center velocity described above, Centaurus-Far got the closest, with a minimum separation of \SI{25}{pc} from the center, while the two Chameleons got as close as \SI{46}{pc} and \SI{40}{pc} to the center in the last \SI{10}{Myr}.
The observation is robust with respect to the precise choice of center velocity, and the minimum separations vary by \SI{10}{pc} or less if we assume the velocity of the center to be the average velocity of all YSO clusters in the vicinity (see \Cref{sec:dynamics}).
Given the large uncertainty in the velocities and ages and the simplistic model, these discrepancies appear modest, and the velocities and ages seem to be in good agreement with the hypothesis of an expanding half-ring.

\begin{figure}[!htbp]
   \centering
   \includegraphics[width=0.95\hsize,keepaspectratio]{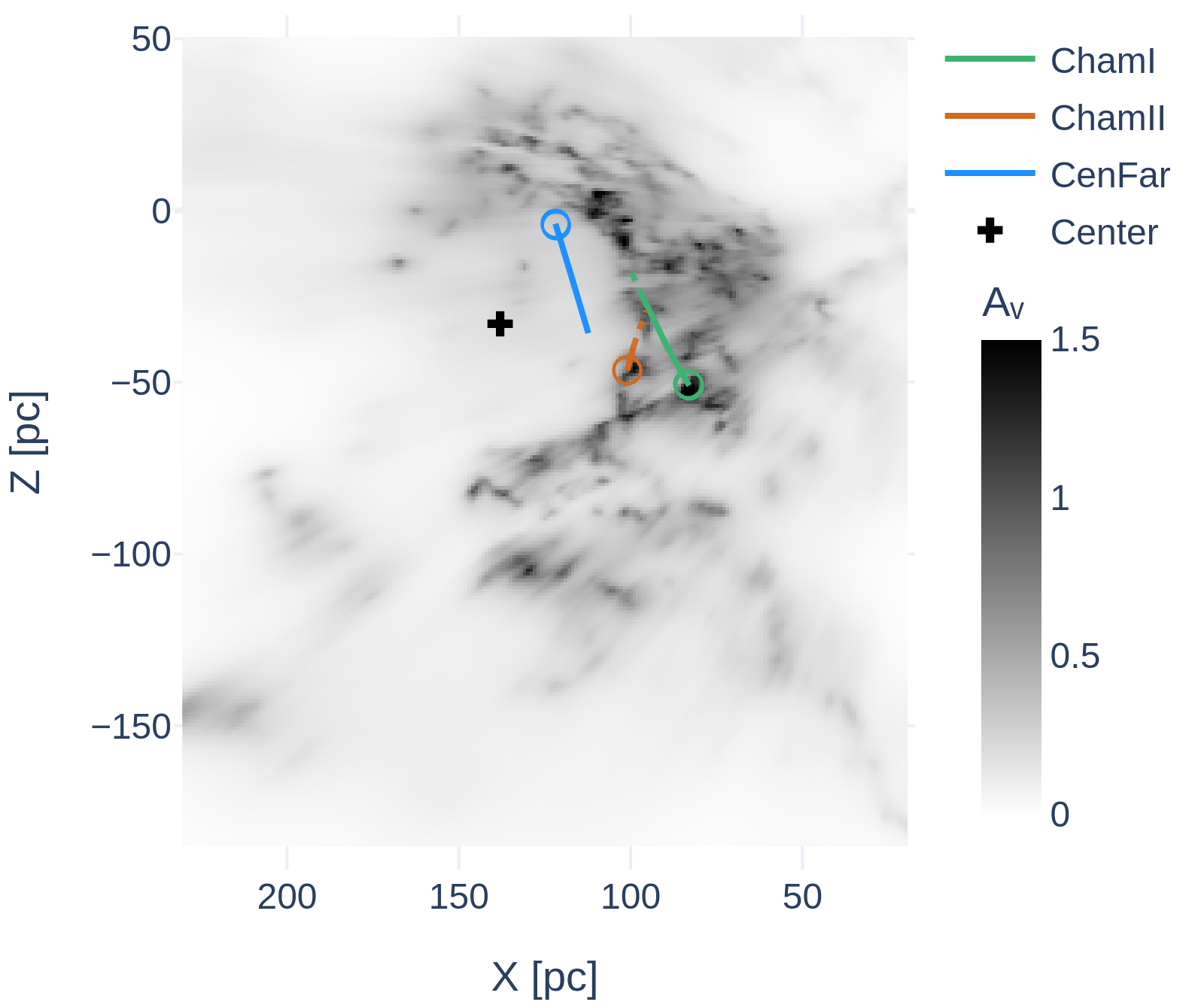}
   \caption{%
      Cartesian X-Z projections of the C.
      The colorbar is linear and clipped at $A_\mathrm{V} = \SI{1.5}{mag}$.
      The traced-back clusters relative to the center are shown as colored lines.
      The solid lines show the traces of the cluster up to their 0.84-age quantile, while the dash lines show the traces back to \SI{-10}{Myr} whenever their 0.84-age quantile is lower.
      See the interactive version of \Cref{fig:edenhofer23_isosurface} and toggle on the clusters to see them in full spatial 3D.
      \label{fig:edenhofer23_xyz}
   }
\end{figure}

\autoref{fig:edenhofer23_xyz} shows the C in Cartesian X-Z projections.
% As the C is almost perfectly perpendicular to the line of sight, the Cartesian X-Y- and Y-Z-projections are not shown.
The three clusters and their trace-backs relative to the center are shown with colored lines.
% Note, the Galactic X axis is flipped as the C lies at negative Galactic Y and would otherwise appear reversed relative to the spherical projection in \Cref{fig:edenhofer23_lb}.

\subsection{Caveats}
\label{sec:caveats}

We have determined the shape and mass of the C with considerable accuracy using 3D interstellar dust maps.
However, the dynamics and energy calculations remain uncertain.
The C's overall shape and the velocities of the YSO clusters suggest it may be an expanding half-ring, but our analysis is constrained by the limited number of YSO clusters observed.
The uncertainties in the velocity measurements are comparable to the relative uncertainties, further complicating the analysis.
In addition to the velocity uncertainties, the cause for the velocities is not known, and they might be purely turbulent in nature.
Finally, although the C surrounds a largely empty space, there is evidence of spurs of interstellar dust at its center, as shown in \autoref{fig:edenhofer23_xyz}.
%However, as the density is low and faint densities have a high distance uncertainty in 3D interstellar dust maps, the faint densities might also be at a different distance, for example, in front or behind the cavity.

In the future, we hope that better velocity measurements will help us quantify the motion of the C much more thoroughly.
However, even much improved velocity uncertainties will not enable us to determine whether they are turbulent or driven by expansion.
Considering the peculiar shape of the C, we believe turbulence alone is unlikely to be the sole cause.

\section{Discussion and conclusion}
\label{sec:discussion_and_conclusion}

Three-dimensional maps of interstellar dust have unlocked new ways to explore the intricate structures within the ISM in comprehensive 3D detail.
Using these maps, we have established that the Chameleon, Musca, and Coalsack molecular clouds are spatially connected, forming a C-shaped structure that encompasses a striking, well-defined cavity.
This structure, which we refer to as the C, is a half-ring structure with a radius of approximately \SI{50}{pc} and a mass of about \SI{5e4}{M_{\odot}}, or \SI{9e4}{M_{\odot}} including everything in the vicinity of the C-shaped cloud.
The C extends roughly \SI{50}{\degree} in both longitude and latitude, and has a depth of about \SI{55}{pc}.
The structure is discernible only through 3D reconstructions of interstellar dust, as obscuration and projection effects obscure it in conventional 2D views.

We propose a single origin for all three molecular clouds comprising the C.
The orbits of clusters in the C are indicative of an expansion that, when combined with the mass of the C, requires the amount of energy released by a single supernova explosion.
We suggest that a singular supernova explosion shaped an existing dense cloud into the C between 4 Myr and 10 Myr ago.

The realization that the Chameleon, Musca, and Coalsack molecular clouds are interconnected in 3D space introduces a novel perspective for their study.
These previously distinct clouds are now understood to be parts of a single large cloud, the C.
This broader context is especially significant for understanding Musca, for which a large body of observational work has been undertaken over the last decade to explain its formation.
Viewing Musca within the scope of the C suggests a relatively simple formation mechanism: cloud formation on an expanding ring, likely driven by stellar feedback.
This insight suggests that feedback-driven expansion is the primary cause for the shape of Musca \citep[see][for the role of magnetic fields]{Inutsuka2015,Ntormousi2017} and calls for a new analysis of the existing data on this cloud.
This new perspective on Musca's formation and shape is a reminder that the environment in which a molecular cloud resides can significantly influence its formation mechanism and morphology.
This work highlights the importance of considering the broader context offered by the new 3D dust maps when interpreting molecular cloud formation and evolution.

\begin{acknowledgements}
    The authors would like to thank Cameren~Swiggum for many helpful comments and discussions throughout the work.
    The authors would like to thank the anonymous referee for their very constructive and fruitful feedback.
    Gordian Edenhofer acknowledges that support for this work was provided by the German Academic Scholarship Foundation in the form of a PhD scholarship (``Promotionsstipendium der Studienstiftung des Deutschen Volkes'').
    This work was co-funded by the European Union (ERC, ISM-FLOW, 101055318).
    Views and opinions expressed are, however, those of the author(s) only and do not necessarily reflect those of the European Union or the European Research Council. Neither the European Union nor the granting authority can be held responsible for them.
\end{acknowledgements}

\FloatBarrier  % Stop figures from being shifted to after the references

\bibliographystyle{aa}  % style aa.bst
\bibliography{literature}

\begin{appendix}
\FloatBarrier

\section{Computing the mass}
\FloatBarrier
\label{appx:mass}

To estimate the mass in a given volume, we first convert the extinction density to a hydrogen density and then integrate the hydrogen density over the given volume.
We adopt the conversion ratio $n_\mathrm{H} = 1653\,\mathrm{cm}^{-3} \rho$ as derived in \citet{ONeill2024LocalChimney} using \citet{Draine2003,Draine2009} to convert our interstellar dust density $\rho$ to a total hydrogen nuclei density $n_\mathrm{H}$.
Analogously to \citet{ONeill2024LocalChimney} we convert the hydrogen nuclei density to a mass via $M = 1.37 \cdot m_p \cdot \sum_i n_{\mathrm{H},i} \cdot \mathrm{d}v_i$, adopting a mean molecular weight of hydrogen ($\mu$) of 1.37, $m_p$ the proton mass, and $\mathrm{d}v_i$ the volume of the $i$th voxel $n_{\mathrm{H},i}$.

\FloatBarrier
\section{Validation of dynamics}
\FloatBarrier
\label{appx:validation_of_dynamics}

In this appendix we study the robustness of the observed relative motion in Galactic Z of the two Chameleon clusters and Centaurus-Far with respect to the center.
The center velocity is defined in the main body of the text via $\nicefrac{1}{2} \cdot \left( v_\mathrm{CenFar} + \nicefrac{1}{2} \cdot \left( v_\mathrm{ChamI} + v_\mathrm{ChamII} \right) \right)$ and equals ${\left(-9.0,-20.0,-8.2\right)}^T\,\si{km/s}$.
Relative to this center velocity, the motion of Centaurus-Far is ${\left(0.8,0.7,2.3\right)}^T\,\si{km/s}$, the motion of Chameleon~I is ${\left(-1.9,0.1,-3.1\right)}^T\,\si{km/s}$, and the motion of Chameleon~II is ${\left(0.4,-1.6,-1.5\right)}^T\,\si{km/s}$.

To test the dependence on the choice of center velocity, we tested an alternative definition of the center velocity based on all nearby YSO clusters.
Specifically, we assumed that the center velocity is the average velocity of all YSO clusters in the catalog of \citet{Hunt2023} within $-20 \leq X \leq 260$\,\si{pc}, $-300 \leq Y \leq -10$\,\si{pc}, and $-240 \leq Z \leq 80$\,\si{pc} and the quality and age cuts from \Cref{sec:dynamics}.
This selects $20$ YSO clusters comprising $2986$ stars mostly in and around the Scorpius–Centaurus association with a median age of \SI{7.3}{Myr} and a maximum 0.84-age-quantile of \SI{18.7}{Myr}.
The center motion defined in this way is ${\left(-8.0,-18.2,-7.2\right)}^T\,\si{km/s}$.
Compared to the center velocity defined via the Chameleon clusters and Centaurus-Far, the center velocity only shifts by \SI{1}{km/s} in Galactic Z while the difference between the motion in Galactic Z of the Chameleons and Centaurus-Far is \SI{5.4}{km/s} respectively \SI{3.8}{km/s}.
Thus, the finding that the two Chameleon clusters are moving down and Centaurus-Far is moving up holds irrespective of the choice of center.

Next, we tested the sensitivity of the cluster velocities with respect to the stellar velocity uncertainties.
Considering the large spatial separation of the two Chameleon clusters and Centaurus-Far, we neglected systematic uncertainties in the clustering itself \citep[cf. the onlinen figure\footnotemark{} from ][]{Ratzenboeck2023ScoCen}.
Instead, we focused on the purely statistical uncertainties of the cluster velocities.
\footnotetext{\url{https://homepage.univie.ac.at/sebastian.ratzenboeck/wp-content/uploads/2024/04/scocen_ages-2.html}}

The velocity uncertainty of the young clusters is dominated by the spread of the stellar velocities.
To estimate the velocity uncertainty of the clusters, we used the cluster assignment from \citet{Ratzenboeck2023}, selecting the high-quality stars in the YSO cluster, and then computed the standard deviation of the sample.
We adopted heliocentric Cartesian velocities based on \textit{Gaia} DR3 proper motion and radial velocity measurements \citep{GaiaCollaboration2022,Katz2023} and added APOGEE DR17 \citep{Apogee2Data2022} and GALAH DR3 \citep{Galah2021} radial velocities when available.
For stars with radial velocity measurements in multiple surveys we used the measurement with the lowest uncertainty.
We selected all stars in a cluster with a radial velocity error $\mathrm{RV}_\mathrm{e} < \SI{10}{km/s}$ and absolute radial velocity $|\mathrm{RV}| < \SI{50}{km/s}$, computed the mean and the standard deviation of the radial velocity of the sample and sub-selected all stars within one standard deviation in radial velocity.
In total, we selected 19 stars for Centaurus-Far, 29 for Chameleon~I, and 3 for Chameleon~II.
Due to the low number of high-quality stars in Chameleon~II, we combine the two Chameleon clusters for the purpose of computing their velocity uncertainty.
The standard deviation of the velocities of this sub-selection for Centaurus-Far is ${\left(2.6, 3.1, 1.0\right)}^T\,\si{km/s}$ and for both Chameleon clusters combined is ${\left(2.3, 3.8, 1.3\right)}^T\,\si{km/s}$.
The retrieved uncertainties in Galactic Z are only slightly below the relative velocity that we find.

\end{appendix}

\end{document}